\documentclass[twocolumn]{aastex631}

\usepackage{amsmath}
\usepackage[encapsulated]{CJK}

\newcommand{\mjysr}{MJy sr$^{-1}$}
\newcommand{\msun}{M$_{\odot}$}

\newcommand{\msolpcsq}{M$_\odot$\,pc$^{-2}$}
\newcommand{\hi}{\mbox{\sc{Hi}$\,$}}
\newcommand{\htwo}{H$_{2}$}

\shorttitle{Diffuse PAH Emission}
\shortauthors{Sandstrom et al.}
\graphicspath{{./}{figures/}}

\begin{document}

\title{PHANGS-JWST First Results: Tracing the Diffuse ISM with JWST Imaging of Polycyclic Aromatic Hydrocarbon Emission in Nearby Galaxies}

\correspondingauthor{Karin Sandstrom}
\email{kmsandstrom@ucsd.edu}

\author[0000-0002-4378-8534]{Karin M. Sandstrom}
\affiliation{Center for Astrophysics \& Space Sciences, Department of Physics, University of California, San Diego, 9500 Gilman Drive, San Diego, CA 92093, USA}

\author[0000-0001-9605-780X]{Eric W. Koch}
\affiliation{Center for Astrophysics $|$ Harvard \& Smithsonian, 60 Garden Street, Cambridge, MA, 02138, USA}

\author[0000-0002-2545-1700]{Adam K. Leroy}
\affiliation{Department of Astronomy, The Ohio State University, 140 West 18th Avenue, Columbus, OH 43210, USA}

\author[0000-0002-5204-2259]{Erik Rosolowsky}
\affiliation{Department of Physics, University of Alberta, Edmonton, Alberta, T6G 2E1, Canada}

\author[0000-0002-6155-7166]{Eric Emsellem}
\affiliation{European Southern Observatory, Karl-Schwarzschild-Stra{\ss}e 2, 85748 Garching, Germany}
\affiliation{Univ Lyon, Univ Lyon1, ENS de Lyon, CNRS, Centre de Recherche Astrophysique de Lyon UMR5574, F-69230 Saint-Genis-Laval France}

\author[0000-0002-0820-1814]{Rowan J. Smith}
\affiliation{Jodrell Bank Centre for Astrophysics, Department of Physics and Astronomy, University of Manchester, Oxford Road, Manchester M13 9PL, UK}

\author[0000-0002-4755-118X]{Oleg V. Egorov}
\affiliation{Astronomisches Rechen-Institut, Zentrum f\"{u}r Astronomie der Universit\"{a}t Heidelberg, M\"{o}nchhofstra\ss e 12-14, 69120 Heidelberg, Germany}

\author[0000-0002-0786-7307]{Thomas G. Williams}
\affiliation{Sub-department of Astrophysics, Department of Physics, University of Oxford, Keble Road, Oxford OX1 3RH, UK}
\affiliation{Max-Planck-Institut f\"{u}r Astronomie, K\"{o}nigstuhl 17, D-69117, Heidelberg, Germany}

\author[0000-0003-3917-6460]{Kirsten L. Larson}
\affiliation{AURA for the European Space Agency (ESA), Space Telescope Science Institute, 3700 San Martin Drive, Baltimore, MD 21218, USA}

\author[0000-0003-0946-6176]{Janice C. Lee}
\affiliation{Gemini Observatory/NSF’s NOIRLab, 950 N. Cherry Avenue, Tucson, AZ, USA}
\affiliation{Steward Observatory, University of Arizona, 933 N Cherry Ave,Tucson, AZ 85721, USA}

\author[0000-0002-3933-7677]{Eva Schinnerer}
\affiliation{Max-Planck-Institut f\"{u}r Astronomie, K\"{o}nigstuhl 17, D-69117, Heidelberg, Germany}

\author[0000-0002-8528-7340]{David A. Thilker}
\affiliation{Department of Physics and Astronomy, The Johns Hopkins University, Baltimore, MD 21218, USA}

\author[0000-0003-0410-4504]{Ashley.~T.~Barnes}
\affiliation{Argelander-Institut f\"{u}r Astronomie, Universit\"{a}t Bonn, Auf dem H\"{u}gel 71, 53121, Bonn, Germany}

\author[0000-0002-2545-5752]{Francesco Belfiore}
\affiliation{INAF — Arcetri Astrophysical Observatory, Largo E. Fermi 5, I-50125, Florence, Italy}

\author[0000-0003-0166-9745]{F. Bigiel}
\affiliation{Argelander-Institut f\"ur Astronomie, Universit\"at Bonn, Auf dem H\"ugel 71, 53121 Bonn, Germany}

\author[0000-0003-4218-3944]{Guillermo A. Blanc}
\affiliation{The Observatories of the Carnegie Institution for Science, 813 Santa Barbara St., Pasadena, CA, USA}
\affiliation{Departamento de Astronom\'{i}a, Universidad de Chile, Camino del Observatorio 1515, Las Condes, Santiago, Chile}

\author[0000-0002-5480-5686]{Alberto D. Bolatto}
\affiliation{Department of Astronomy and Joint Space-Science Institute, University of Maryland, College Park, MD 20742, USA}

\author[0000-0003-0946-6176]{Médéric~Boquien}
\affiliation{Centro de Astronomía (CITEVA), Universidad de Antofagasta, Avenida Angamos 601, Antofagasta, Chile}

\author[0000-0001-5301-1326]{Yixian Cao}
\affiliation{Max-Planck-Institut f\"ur Extraterrestrische Physik (MPE), Giessenbachstr. 1, D-85748 Garching, Germany}

\author[0000-0002-5235-5589]{J\'er\'emy Chastenet}
\affil{Sterrenkundig Observatorium, Universiteit Gent, Krijgslaan 281 S9, B-9000 Gent, Belgium}

\author[0000-0002-5635-5180]{M\'elanie Chevance}
\affiliation{Institut f\"{u}r Theoretische Astrophysik, Zentrum f\"{u}r Astronomie der Universit\"{a}t Heidelberg,\\ Albert-Ueberle-Strasse 2, 69120 Heidelberg, Germany}
\affiliation{Cosmic Origins Of Life (COOL) Research DAO, coolresearch.io}

\author[0000-0003-2551-7148]{I-Da Chiang \begin{CJK*}{UTF8}{bkai}(江宜達)\end{CJK*}}%
\affiliation{Institute of Astronomy and Astrophysics, Academia Sinica, No. 1, Sec. 4, Roosevelt Road, Taipei 10617, Taiwan}

\author[0000-0002-5782-9093]{Daniel~A.~Dale}
\affiliation{Department of Physics and Astronomy, University of Wyoming, Laramie, WY 82071, USA}

\author[0000-0001-5310-467X]{Christopher M. Faesi}
\affiliation{University of Connecticut, Department of Physics, 196A  Auditorium Road, Unit 3046, Storrs, CT, 06269}

\author[0000-0001-6708-1317]{Simon C.~O.\ Glover}
\affiliation{Universit\"{a}t Heidelberg, Zentrum f\"{u}r Astronomie, Institut f\"{u}r Theoretische Astrophysik, Albert-Ueberle-Stra{\ss}e 2, D-69120 Heidelberg, Germany}

\author[0000-0002-3247-5321]{Kathryn~Grasha}
\affiliation{Research School of Astronomy and Astrophysics, Australian National University, Canberra, ACT 2611, Australia}   
\affiliation{ARC Centre of Excellence for All Sky Astrophysics in 3 Dimensions (ASTRO 3D), Australia} 

\author[0000-0002-9768-0246]{Brent Groves}
\affiliation{International Centre for Radio Astronomy Research, University of Western Australia, 7 Fairway, Crawley, 6009 WA, Australia}

\author[0000-0002-8806-6308]{Hamid Hassani}
\affiliation{Department of Physics, University of Alberta, Edmonton, Alberta, T6G 2E1, Canada}

\author[0000-0001-9656-7682]{Jonathan~D.~Henshaw}
\affiliation{Astrophysics Research Institute, Liverpool John Moores University, 146 Brownlow Hill, Liverpool L3 5RF, UK}
\affiliation{Max-Planck-Institut f\"ur Astronomie, K\"onigstuhl 17, D-69117 Heidelberg, Germany}

\author[0000-0002-9181-1161]{Annie~Hughes}
\affiliation{IRAP, Universit\'e de Toulouse, CNRS, CNES, UPS, (Toulouse), France} 

\author[0000-0002-0432-6847]{Jaeyeon Kim}
\affiliation{Zentrum f\"{u}r Astronomie der Universit\"{a}t Heidelberg, Institut f\"{u}r Theoretische Astrophysik, Albert-Ueberle-Str. 2, 69120 Heidelberg}

\author[0000-0002-0560-3172]{Ralf S.\ Klessen}
\affiliation{Universit\"{a}t Heidelberg, Zentrum f\"{u}r Astronomie, Institut f\"{u}r Theoretische Astrophysik, Albert-Ueberle-Stra{\ss}e 2, D-69120 Heidelberg, Germany}
\affiliation{Universit\"{a}t Heidelberg, Interdisziplin\"{a}res Zentrum f\"{u}r Wissenschaftliches Rechnen, Im Neuenheimer Feld 205, D-69120 Heidelberg, Germany}

\author[0000-0001-6551-3091]{Kathryn Kreckel}
\affiliation{Astronomisches Rechen-Institut, Zentrum f\"{u}r Astronomie der Universit\"{a}t Heidelberg, M\"{o}nchhofstra\ss e 12-14, 69120 Heidelberg, Germany}

\author[0000-0002-8804-0212]{J.~M.~Diederik~Kruijssen}
\affiliation{Cosmic Origins Of Life (COOL) Research DAO, coolresearch.io}

\author[0000-0002-1790-3148]{Laura A. Lopez}
\affiliation{Department of Astronomy, The Ohio State University, 140 West 18th Avenue, Columbus, Ohio 43210, USA}
\affiliation{Center for Cosmology and Astroparticle Physics, 191 West Woodruff Avenue, Columbus, OH 43210, USA}
\affiliation{Flatiron Institute, Center for Computational Astrophysics, NY 10010, USA}

\author[0000-0001-9773-7479]{Daizhong Liu}
\affiliation{Max-Planck-Institut f\"ur Extraterrestrische Physik (MPE), Giessenbachstr. 1, D-85748 Garching, Germany}

\author[0000-0002-6118-4048]{Sharon E. Meidt}
\affiliation{Sterrenkundig Observatorium, Universiteit Gent, Krijgslaan 281 S9, B-9000 Gent, Belgium}

\author[0000-0001-7089-7325]{Eric J.\,Murphy}
\affiliation{National Radio Astronomy Observatory, 520 Edgemont Road, Charlottesville, VA 22903, USA}

\author[0000-0002-1370-6964]{Hsi-An Pan}
\affiliation{Department of Physics, Tamkang University, No.151, Yingzhuan Road, Tamsui District, New Taipei City 251301, Taiwan} 

\author[0000-0002-0472-1011]{Miguel~Querejeta}
\affiliation{Observatorio Astron\'{o}mico Nacional (IGN), C/Alfonso XII, 3, E-28014 Madrid, Spain}

\author[0000-0002-2501-9328]{Toshiki Saito}
\affiliation{National Astronomical Observatory of Japan, 2-21-1 Osawa, Mitaka, Tokyo, 181-8588, Japan}

\author[0000-0002-5783-145X]{Amy Sardone}
\affiliation{Department of Astronomy, The Ohio State University, 140 West 18th Avenue, Columbus, OH 43210, USA}
\affiliation{Center for Cosmology and Astroparticle Physics, 191 West Woodruff Avenue, Columbus, OH 43210, USA}

\author[0000-0001-6113-6241]{Mattia C. Sormani}
\affiliation{Universit\"{a}t Heidelberg, Zentrum f\"{u}r Astronomie, Institut f\"{u}r Theoretische Astrophysik, Albert-Ueberle-Str 2, D-69120 Heidelberg, Germany}

\author[0000-0002-9183-8102]{Jessica Sutter}
\affiliation{Center for Astrophysics \& Space Sciences, Department of Physics, University of California, San Diego, 9500 Gilman Drive, San Diego, CA 92093, USA}

\author[0000-0003-1242-505X]{Antonio Usero}
\affiliation{Observatorio Astron\'{o}mico Nacional (IGN), C/Alfonso XII, 3, E-28014 Madrid, Spain}

\author[0000-0002-7365-5791]{Elizabeth~J.~Watkins}
\affiliation{Astronomisches Rechen-Institut, Zentrum f\"{u}r Astronomie der Universit\"{a}t Heidelberg, M\"{o}nchhofstra\ss e 12-14, 69120 Heidelberg, Germany}

\suppressAffiliations

\begin{abstract}

JWST observations of polycyclic aromatic hydrocarbon (PAH) emission provide some of the deepest and highest resolution views of the cold interstellar medium (ISM) in nearby galaxies. If PAHs are well mixed with the atomic and molecular gas and illuminated by the average diffuse interstellar radiation field, PAH emission may provide an approximately linear, high resolution, high sensitivity tracer of diffuse gas surface density. We present a pilot study that explores using PAH emission in this way based on MIRI observations of IC~5332, NGC~628, NGC~1365, and NGC~7496 from the PHANGS-JWST Treasury. Using scaling relationships calibrated in \citet{LEROY1_PHANGSJWST}, scaled F1130W provides 10--40 pc resolution and 3$\sigma$ sensitivity of $\Sigma_{\rm gas} \sim 2$ \msun\ pc$^{-2}$. We characterize the surface densities of structures seen at $<7$ \msun\ pc$^{-2}$ in our targets, where we expect the gas to be HI-dominated. We highlight the existence of filaments, inter-arm emission, and holes in the diffuse ISM at these low surface densities. Below $\sim 10$ \msun\ pc$^{-2}$ for NGC~628, NGC~1365, and NGC~7496 the gas distribution shows a ``Swiss cheese’’-like topology due to holes and bubbles pervading the relatively smooth distribution of diffuse ISM. Comparing to recent galaxy simulations, we observe similar topology for the low surface density gas, though with notable variations between simulations with different setups and resolution. Such a comparison of high resolution, low surface density gas with simulations is not possible with existing atomic and molecular gas maps, highlighting the unique power of JWST maps of PAH emission.

\end{abstract}


\received{October 24, 2022}
\accepted{December 6, 2022}
\submitjournal{The Astrophysical Journal Letters}

\section{Introduction} \label{sec:intro}

With currently available radio telescopes, observations of galaxies outside the Local Group do not have sufficient resolution and sensitivity to characterize the structure and column densities of the atomic gas dominated diffuse interstellar medium (ISM) on scales of less than hundreds of parsecs. Observations in the Milky Way show that at small spatial scales this phase of the ISM has structure and characteristics that reflect the interaction of turbulence, feedback, galactic dynamics, magnetic fields, gas phase transitions, and more \citep{ehlerov2005,kalberla2009,suad2014,li2021,soler2022}. The insights provided by studying the diffuse ISM in the Milky Way and Local Group at high resolution inform our understanding of key processes in ISM physics, but are limited by distance ambiguities in the Milky Way and the mainly non-representative sample of galaxies available in the Local Group (i.e.\ a nearly quiescent massive spiral, two dwarf-spirals, and several low mass dwarf irregulars). It is therefore of great interest to characterize the properties of the diffuse ISM at $\lesssim 100$ pc resolution in other massive star-forming galaxies, but until the Square Kilometer Array or the Next Generation VLA are operational, this is not possible directly with \hi\ 21 cm observations.

One approach to studying the diffuse ISM is to use dust as a tracer of gas. With a known dust-to-gas ratio (D/G), assumed to apply across ISM phases, a measured dust surface density can be converted to a gas surface density \citep[e.g.][]{bot2007,draine2007b,peretto2009,leroy2011,eales2012,scoville2014,romanduval2014,groves2015}. Such a technique does not generally overcome the resolution limitations, however, since most tracers of the dust surface density are observed at coarse resolution. For instance, far-IR images of galaxies from the Herschel Space Observatory ($\sim11''$ at 160 \micron) still have resolution lower than what can be achieved with VLA \hi\ observations.   

Observations of polycyclic aromatic hydrocarbon (PAH) emission from JWST open up new pathways to probe the diffuse ISM at $\lesssim100$ pc scales in nearby galaxies because of their high angular resolution and sensitivity. Tracing gas with PAH emission should be possible if: 1) dust and PAHs are mixed with gas with approximately constant ratios of dust-to-gas and PAH-to-dust \citep[a.k.a.\ ``PAH fraction''][]{draine2007a} and 2) the PAHs are illuminated by an approximately constant or known radiation field intensity and spectrum. Because PAHs experience single-photon heating over a wide range of radiation field intensities \citep{li2001}, their resulting emission intensity varies linearly with the incident radiation field. If that radiation field is known (or can be calibrated locally), the PAH emission then linearly traces the PAH surface density, which could then be related to gas mass with a PAH-to-gas ratio, as described above. Variations in the PAH fraction and PAH-to-gas ratio are active areas of research \citep[including][in this Issue]{LEROY1_PHANGSJWST,CHASTENET1_PHANGSJWST,EGOROV_PHANGSJWST}. Outside the vicinity of {\sc H~II} regions, PAH abundance appears to be relatively constant in the cold atomic and molecular ISM, though dependent on metallicity \citep{CHASTENET1_PHANGSJWST}. Future work using JWST observations will be critical to characterizing the PAH fraction across a variety of galactic environments. 

For the radiation field, the diffuse ISM in particular is expected to be dominated by the average diffuse interstellar radiation field (ISRF), which we will refer to as the diffuse ISRF.  In the Milky Way, the diffuse ISRF is found to be powered predominantly by B stars \citep{mathis83} and does not vary strongly on small scales (though it does change radially in the Galaxy). The diffuse ISRF dominates dust heating outside the sphere of influence of star-forming regions \citep[$\sim$100 pc in the Magellanic Clouds;][]{lawton2010} and outside the shielding interiors of dense clouds. By isolating PAH emission in regions where dust heating is dominated by the diffuse ISRF we can hope to construct a tracer for gas column based on the PAH emission. 

Previous generations of infrared telescopes did not have the resolution to separate out the diffuse ISRF-powered PAH emission spatially outside the Local Group. At the typical $\sim100{-}1000$ pc spatial resolution of {\em Spitzer} in nearby galaxies ($D\sim10$ Mpc), the sphere of influence of individual star-forming regions is not resolved, leaving the measurement of the fraction of PAH emission being powered by the diffuse ISRF reliant on modeling.  However, infrared SED modeling of the mid- to far-IR emission from galaxies suggests that a substantial fraction of the dust and PAH emission in galaxies arises from diffuse ISRF heating \citep[e.g.][]{draine2007b,aniano12,galliano2018,aniano20}.

The idea of tracing diffuse ISM gas with PAH emission has a long heritage from studies of the high latitude ``cirrus'' emission in the Milky Way. Observations with IRAS found that the PAH-dominated broadband emission at 12 \micron\ was well correlated with \hi\ at high Galactic latitudes \citep{boulanger1985,puget1985,boulanger1988}. Observations with the AROME balloon experiment mapped the 3.3 \micron\ PAH feature specifically and found it to be well correlated with the 12 \micron\ emission from IRAS and \hi\ \citep{girard1988,girard1994}. COBE/DIRBE observations from 3.5-240 \micron\ allowed measurement of the mean spectrum of the \hi-correlated high latitude ``cirrus'' dust emission, which includes prominent PAH features \citep{dwek1997}. Using {\em Spitzer} and {\em Herschel} observations, \citet{compiegne2010} found that IRAC 8 \micron\ emission linearly scaled with a combination of PAH abundance, ISM column density, and the intensity of the ISRF, as would be expected for grains undergoing single-photon heating. At this point, the association of PAH emission with diffuse \hi\ at high latitudes in the Milky Way is well enough understood that it is a key constraint on essentially all dust models \citep[see recent summary in][]{hensley2022}. 

In this Letter, we present a case study for using PAH emission observed with JWST to trace low surface density gas at 10$-$40 pc resolution in nearby galaxies.  We show maps of the diffuse ISM at $\Sigma_{\rm gas} < 7$ \msolpcsq, where we expect the gas to be primarily atomic \citep{KrumholzMcKee2009ApJ...693..216K,SternbergLePetit2014ApJ...790...10S,BialySternberg2016ApJ...822...83B}. We identify low gas surface density structures in interarm regions, bubbles, and the outskirts of galaxies that are detected at high signal-to-noise and resolution in the JWST maps. While for this study we extrapolate a simple scaling determined in the H$_2$-dominated regions of the galaxy, in future work, more detailed modeling will be needed to address the uncertainties inherent in the PAH emission to gas surface density conversion. It is clear, however, from these early JWST observations, that the gas surface density sensitivity of MIRI observations in nearby galaxies far exceeds that of existing mm and radio observations. 

We proceed as follows: in Section~\ref{sec:obs} we briefly describe the JWST-MIRI observations.  We describe using PAH emission as a gas tracer in Section~\ref{sec:pahtogas}. In Section~\ref{sec:results} we present maps of the gas surface in our target galaxies. We characterize the topology of the diffuse ISM using the {\em genus} statistic \citep{Coles1988MNRAS.234..509C,Koch2019} in Section~\ref{sec:genus}.  In Section~\ref{sec:disc} we discuss the implications of these findings for PAH emission as a gas tracer and highlight the utility of the extremely high resolution view of the atomic ISM available from JWST that will not be equaled before the Next Generation VLA (ngVLA) or Square Kilometer Array (SKA) come online. 

\section{Observations} \label{sec:obs}

We observed NGC~628, NGC~1365, NGC~7496, and IC~5332 with the MIRI imager between July 6 to August 13, 2022 using the F770W, F1000W, F1130W, and F2100W filters. Table~\ref{tab:galaxies} lists relevant properties of our targets. F770W and F1130W are both centered specifically on strong PAH features at 7.7 and 11.3 \micron. Because of the relatively narrow widths of these filters ($\Delta \lambda \sim2.2$ and 0.7 \micron\ for F770W and F1130W, respectively) which are comparable or smaller than the widths of the 7.7 and 11.3 \micron\ PAH features themselves, the fluxes in our targets are generally dominated by PAHs for typical radiation field intensities, with only a minimal contribution from the underlying hot dust continuum \citep[see models from][for example]{draine2007a}. Stellar continuum is only an important contribution to the F770W and F1130W bands towards specific mid-IR bright stars or low-ISM, high stellar surface density regions like the nucleus of NGC~628 \citep{HOYER_PHANGSJWST}, which do not make up a significant fraction of the sight-lines in our maps. Given the dominance of PAH emission at the F770W and F1130W filters in our targets, we do not attempt any removal of the hot dust continuum using F1000W (which may also include a strong or dominant contribution from PAH emission) or F2100W.   

The observations and data reduction for the PHANGS MIRI images are described in \citet{LEE_PHANGSJWST}. Validation of the background levels of the images, which is of particular importance for studying faint emission, follows the methods described in Appendix A and B of \citet{LEROY1_PHANGSJWST}.  Noise levels in the native resolution F770W and F1130W maps are $\sim0.1-0.2$ \mjysr.

\begin{deluxetable*}{lccccccc}[t!]
\tabletypesize{\small}
\tablecaption{Galaxy Properties \label{tab:galaxies}}
\tablewidth{0pt}
\tablehead{
\colhead{Target} & 
\colhead{Distance} &
\colhead{$\log M_*$} &
\colhead{SFR$_{\rm tot}$} &
\colhead{12$+\log$(O/H)} &
\colhead{$i$} &
\colhead{R$_e$} &
\colhead{$\frac{F770W}{I_{\rm CO,2-1}}$, $\frac{F1130W}{I_{\rm CO,2-1}}$} \\
\colhead{ } & 
\colhead{(Mpc)} &
\colhead{(M$_{\odot}$)} &
\colhead{(M$_{\odot}$ yr$^{-1}$)} &
\colhead{(dex)} &
\colhead{($^{\circ}$)} &
\colhead{(kpc)} & 
\colhead{(${\rm MJy \ sr}^{-1} \ ({\rm K \ km~s^{-1}})^{-1}$)}
}
\startdata
IC~5332  &  9.0 &  9.7 & 0.4 & 8.30 & 27 & 3.6 & 0.79, 1.07 \\ 
NGC~628  &  9.8 & 10.3 & 1.7 & 8.53 &  9 & 3.9 & 0.93, 1.29 \\  
NGC~1365 & 19.6 & 11.0 & 17  & 8.67 & 55 & 2.8 & 0.72, 1.07 \\
NGC~7496 & 18.7 & 10.0 & 2.2 & 8.51 & 36 & 3.8 & 1.29, 1.78
\enddata
\tablecomments{Properties adopted following \citet{LEE_PHANGSJWST}, which draws distances from \citet{anand21}; stellar masses ($\log M_*$), star formation rates (SFR$_{\rm tot}$), and effective radius (R$_e$) values from \citet{leroy2021}; inclinations ($i$) from \citet{lang20}; and metallicities ($12+\log$(O/H)) from B. Groves, K. Kreckel et al. (MNRAS submitted). Gas phase metallicities are presented on the $S$-cal system \citep{scal16} estimated at the effective radius $R_e$. The $\frac{F770W}{I_{\rm CO,2-1}}$ and $\frac{F1130W}{I_{\rm CO,2-1}}$ values are taken from \citet{LEROY1_PHANGSJWST}.}
\end{deluxetable*}

\section{Tracing the Diffuse ISM with PAH Emission} \label{sec:pahtogas}

\subsection{Our Approach and Caveats}

As discussed above, tracing gas surface density with PAH emission involves several assumptions: 1) a PAH-to-gas ratio (which can be separated into dust-to-gas and PAH fraction components), and 2) a radiation field intensity and spectrum (in the regime where PAHs experience single-photon heating so the scaling with radiation field intensity is linear).  In this Letter, we make two main assumptions that allow us use PAH emission to estimate the gas column density. These assumptions are likely too simplistic and should be greatly improved in future work, but they let us perform a pilot study for this approach to tracing gas. 

First, we assume that the PAH emission to gas surface density ratio does not vary with cold gas phase. We can then pick regions where we know the gas surface density to empirically estimate this ratio. For this purpose, we select regions that are likely to be H$_2$-dominated where we also have an estimate of the H$_2$ surface density based on ALMA $^{12}$CO $J=(2-1)$ maps. To do this, we convert the CO emission to gas surface density with an assumed CO-to-H$_2$ conversion factor and find the ratio with PAH emission. 

The second assumption we make is that the diffuse ISRF responsible for heating the dust is approximately constant outside the vicinity of star-forming regions. This allows us to apply the PAH emission to gas surface density ratio measured for the molecular gas to the rest of the ISM.  

In practice, we adopt the median ratio of PAH emission to CO~(2-1) emission measured by \citet{LEROY1_PHANGSJWST} for moderate mid-IR intensities $\sim 0.5$ to $30$ \mjysr and $1.7'' \approx 70{-}160$~pc resolution in each target. Over this range of mid-IR intensities, the PAH tracing F770W and F1130W emission correlates strongly with CO and an approximately linear relationship relates the PAH emission to CO emission\footnote{The relationship between CO emission and PAH emission is not exactly linear. We adopt the median ratio for simplicity. If we had chosen to evaluation the power law fits of CO to PAH emission in \citet{LEROY1_PHANGSJWST} at $\sim 1$~\mjysr\ instead, the ratio of PAH emission to CO emission would be $\sim 20{-}50\%$ higher. However, at this low surface brightness, \textsc{Hi} could already make important contributions to the total gas surface density.}. These PAH emission to CO ratios resemble those expected for typical diffuse ISRF values in inner galaxy regions ($U\sim$ few $U_{\rm MMP}$\footnote{$U_{\rm MMP}$ is the ISRF intensity normalized to the \citet{mathis83} Solar neighborhood ISRF.}), PAH fractions ($q_{\rm PAH}\sim 4-5$\%), and dust-to-gas ratios (D/G$\sim 0.01$; see their Equations 4 and 6). But we emphasize that the value we adopt is fundamentally empirical and anchored to the ALMA CO~(2-1) data, not derived from models.

At the high resolution of our data we expect this median PAH to CO emission ratio to capture mostly diffuse ISM emission. Massive star forming regions with high radiation fields are certainly present and associated with bright PAH and continuum emission \citep[e.g.,][]{CHASTENET1_PHANGSJWST,EGOROV_PHANGSJWST,HASSANI_PHANGSJWST}, but as Figures \ref{fig:contour_ic5332} through \ref{fig:contour_7496} show, such bright regions do not make up most of the area in our maps, and typically only dominate the dust heating on scales of $\sim100$ pc \citep[e.g.,][]{lawton2010}.

Adopting these two assumptions and this empirical ratio of PAH emission to CO emission, we calibrate a relationship between PAH emission and gas surface density. Then we extrapolate this scaling to the diffuse ISM, where we expect the gas to be predominantly atomic. We note that we do not mask star forming regions in our resulting maps, but our focus on low surface density gas (or equivalently, low surface brightness PAH emission) means such regions are generally not included in the diffuse gas we consider.

For the purpose of this pilot study, these simple assumptions and this fundamentally empirically approach allow a straightforward scaling of PAH emission to gas surface density, but a number of potential effects will need to be addressed in future work to rigorously calibrate PAH emission as a tracer. There is evidence from measurements of depletion that dust-to-gas ratios vary with ISM density \citep{jenkins09,romanduval2021} by a factor of up to 2, though this depends on whether the observed depleted oxygen is incorporated into dust \citep{whittet2010,choban2022}. Changes in the dust-to-gas ratio between ISM phases at fixed metallicity have also been observed using far-IR, \hi, and CO observations \citep{romanduval2014,chiang2018,vilchez2019}. From high spatial resolution studies of Local Group galaxies \citep{chastenet2019} and the early PHANGS-JWST targets \citep{CHASTENET1_PHANGSJWST,EGOROV_PHANGSJWST}, the PAH fraction appears to be relatively constant in the atomic and molecular phases, but does drop sharply in {\sc H~II} regions, and has a metallicity dependence. Milky Way observations have suggested that PAH abundance can vary between the warm neutral medium (WNM) and cold neutral medium (CNM) \citep{hensley2022}. The diffuse ISRF in galaxies, judged from the $U_{\rm min}$ component in far-IR spectral energy distribution modeling \citep[e.g.][]{draine2007b,aniano12}, appears to vary smoothly and primarily radially in highly resolved Local Group galaxies \citep{draine2014,chastenet2019}. In the Milky Way, the diffuse ISRF varies primarily radially and is dominated by B stars \citep{mathis83}. To summarize, we expect there are a number of effects that can alter the PAH emission to gas surface density conversion, not all of which will be mitigated by our local calibration of the scaling in each galaxy.

\subsection{PAH Emission to Gas Surface Density Scaling}

We adopt PAH emission to CO ratios for each galaxy from \citet{LEROY1_PHANGSJWST}. We note that these values are specific to the four galaxies we target and could change if D/G, $q_{\rm PAH}$, or the diffuse ISRF intensities vary. The average value across all galaxies for the ratio of PAH emission to CO (2$-$1) integrated intensity are:
\begin{equation}\label{eq:ratio77}
    \frac{F770W}{{\rm MJy \ sr}^{-1}} = 0.91 \ \frac{I_{\rm CO,2-1}}{{\rm K \ km~s^{-1}}}
\end{equation}
\begin{equation}\label{eq:ratio113}
    \frac{F1130W}{{\rm MJy \ sr}^{-1}} = 1.26 \ \frac{I_{\rm CO,2-1}}{{\rm K \ km~s^{-1}}}
\end{equation}
These ratios vary slightly between galaxies. The ratio for F770W is 0.79, 0.93, 0.72, and 1.29 in IC~5332, NGC~628, 1365, and 7496, respectively.  For F1130W, it ranges between 1.07, 1.29, 1.07, and 1.78 for the same galaxies \citep[see][Table~3, and Table~\ref{tab:galaxies} above]{LEROY1_PHANGSJWST}.
Assuming the Milky Way CO-to-H$_2$ conversion factor, including He and metals, $\alpha_{\rm CO 1-0}^{\rm MW} = 4.35$ \msun\ pc$^{-2}$ (K km/s)$^{-1}$) \citep{bolatto2013}, and a CO J$=$(2-1) to (1-0) line ratio of $R_{21}=0.65$ \citep{denbrok2021,leroy2022}, we can convert these ratios into PAH emission per gas surface density, assuming all gas is molecular.  These values, for the average ratios shown in Equation~\ref{eq:ratio77} and \ref{eq:ratio113}, are:
\begin{equation}
   \frac{\Sigma_{\rm gas}}{{\rm M_{\odot}} \ {\rm pc}^2}  = 7.4 \times \left(\frac{F770W}{{\rm MJy \ sr}^{-1}}\right) \left( \frac{\alpha_{\rm CO 1-0}}{\alpha_{\rm CO 1-0}^{\rm MW}} \right) \left( \frac{0.65}{R_{21}} \right)
\end{equation}
\begin{equation}\label{eq:sdscalef1130w}
 \frac{\Sigma_{\rm gas}}{{\rm M_{\odot}} \ {\rm pc}^2}  = 5.3 \times \left(\frac{F1130W}{{\rm MJy \ sr}^{-1}}\right) \left( \frac{\alpha_{\rm CO 1-0}}{\alpha_{\rm CO 1-0}^{\rm MW}} \right) \left( \frac{0.65}{R_{21}} \right) 
\end{equation}
Given that the noise levels are comparable in the F770W and F1130W maps at native resolution (between $0.1{-}0.2$ MJy sr$^{-1}$), it is clear that our most sensitive tracer is F1130W, so we proceed in the following to use that as our primary observable. F1130W's point spread function has a FWHM of $0.36''$, yielding 15.7, 17.1, 34.2, and 32.6~pc resolution at the distances of IC~5332, NGC~628, NGC~1365, and NGC~7496, respectively.  We note that at this resolution, which resolves giant molecular cloud complexes and is significantly smaller than the expected scale height of the gas disk, correcting for the inclination of the galaxy introduces an additional assumption about the extent of the emitting structures along the line of sight. The appropriate inclination correction may differ between clouds or a diffuse medium. Therefore, we proceed in the following without an inclination correction, but note that if one were applied, the largest effect would be for NGC~1365 which has an inclination of $i=55^{\circ}$. 

For this pilot study, we use a single scaling factor to convert PAH emission to gas surface density. Several of the components in the conversion of PAH emission to gas surface density are likely to change radially in galaxies. The CO-to-H$_2$ conversion factor likely increases with radius, tracking decreasing metallicity \citep{wolfire2010,leroy2011,glover2012,bolatto2013,madden2020} and also can decrease in the centers of galaxies \citep{sandstrom2013}, likely due to excitation and optical depth effects \citep{teng2022}. Dust-to-gas ratios and PAH fractions may decrease with metallicity, and therefore radius \citep[we note that the observed PAH fraction in our targets does vary with metallicity, but at a relatively minor level given the $\sim Z_{\odot}$ metallicities;][]{CHASTENET1_PHANGSJWST}. The diffuse ISRF also is likely to vary radially, tracing a combination of the stellar mass surface density and star formation rate surface density of the galaxies \citep{mathis83,wolfire2003}.  In future work, we aim to calibrate a radially dependent PAH-to-gas surface density tracer by finding diffuse ISRF dominated, CO-bright regions to measure the scaling factor as a function of radius. We note that some of the factors will likely vary with opposing radial trends, potentially leading to a less variable PAH-to-gas conversion than the individual components.

\section{Results}\label{sec:results}

\begin{figure*}
    \centering
    \includegraphics[height=12cm]{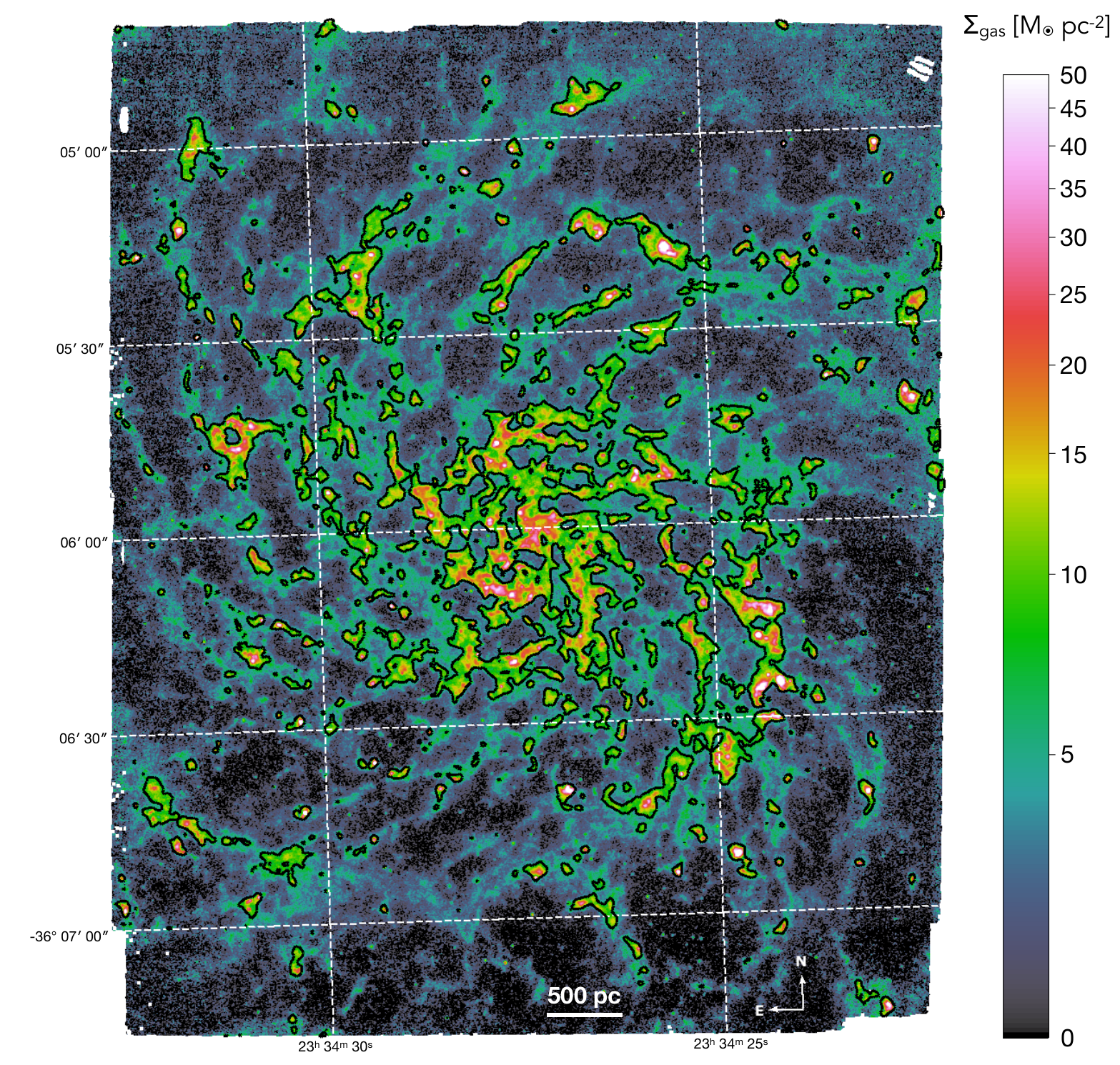}
    \caption{IC~5332 gas surface density from scaled F1130W emission. A single black contour highlights the brightness level where the F1130W emission implies $\Sigma_{\rm gas} = 7$ \msun\ pc$^{-2}$. This figure and the subsequent figures for NGC~628, 1365, and 7496 (Figures~\ref{fig:contour_ngc628}, ~\ref{fig:contour_1365}, and ~\ref{fig:contour_7496}) are all shown with the same color bar. We show a 500 pc scale bar for reference.}
    \label{fig:contour_ic5332}
\end{figure*}

\begin{figure*}
    \centering
    \includegraphics[height=12cm]{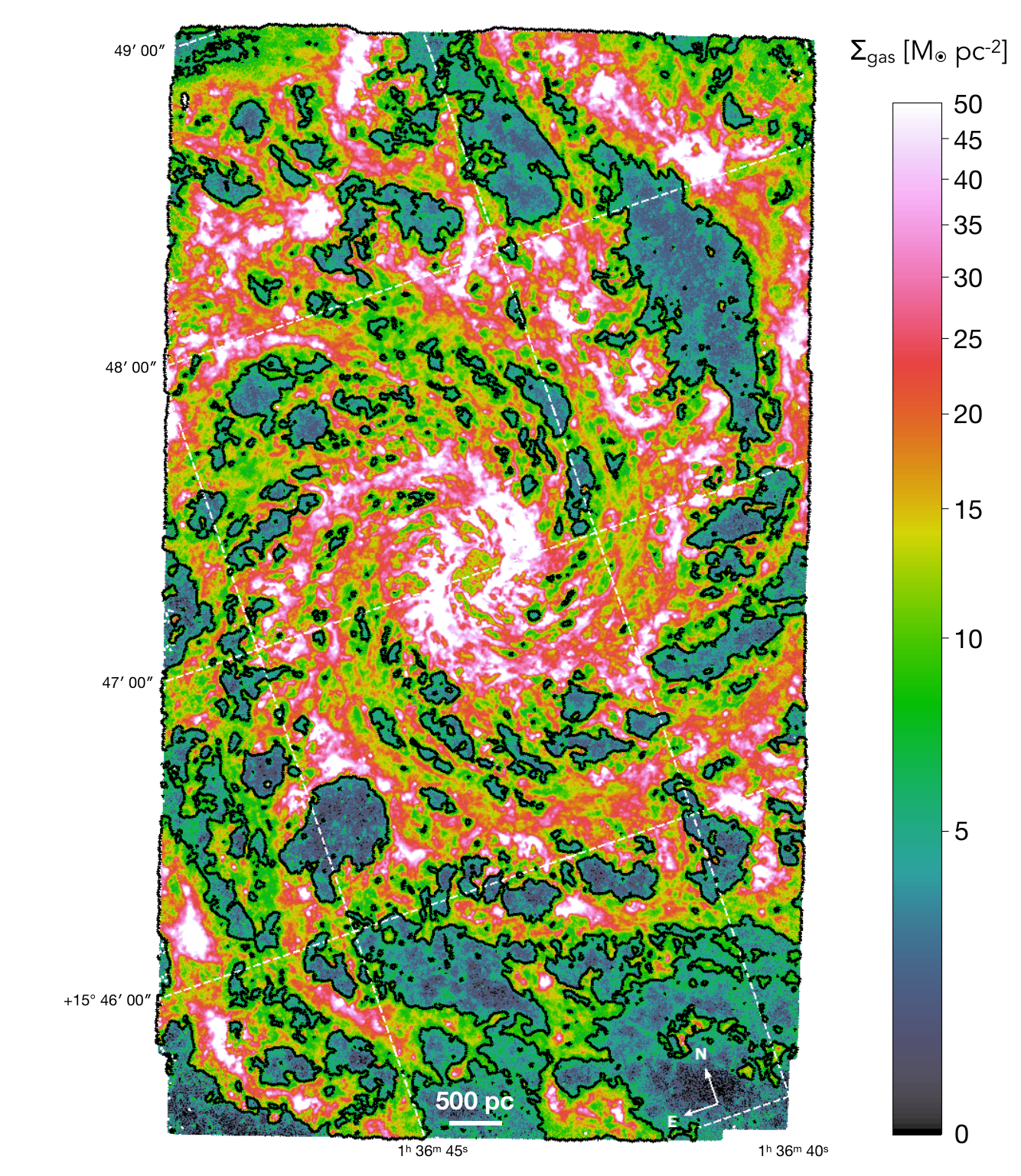}
    \caption{NGC~628 gas surface density from scaled F1130W emission. A single black contour highlights the brightness level where the F1130W emission implies $\Sigma_{\rm gas} = 7$ \msun\ pc$^{-2}$.}
    \label{fig:contour_ngc628}
\end{figure*}

\begin{figure*}
    \centering
    \includegraphics[height=12cm]{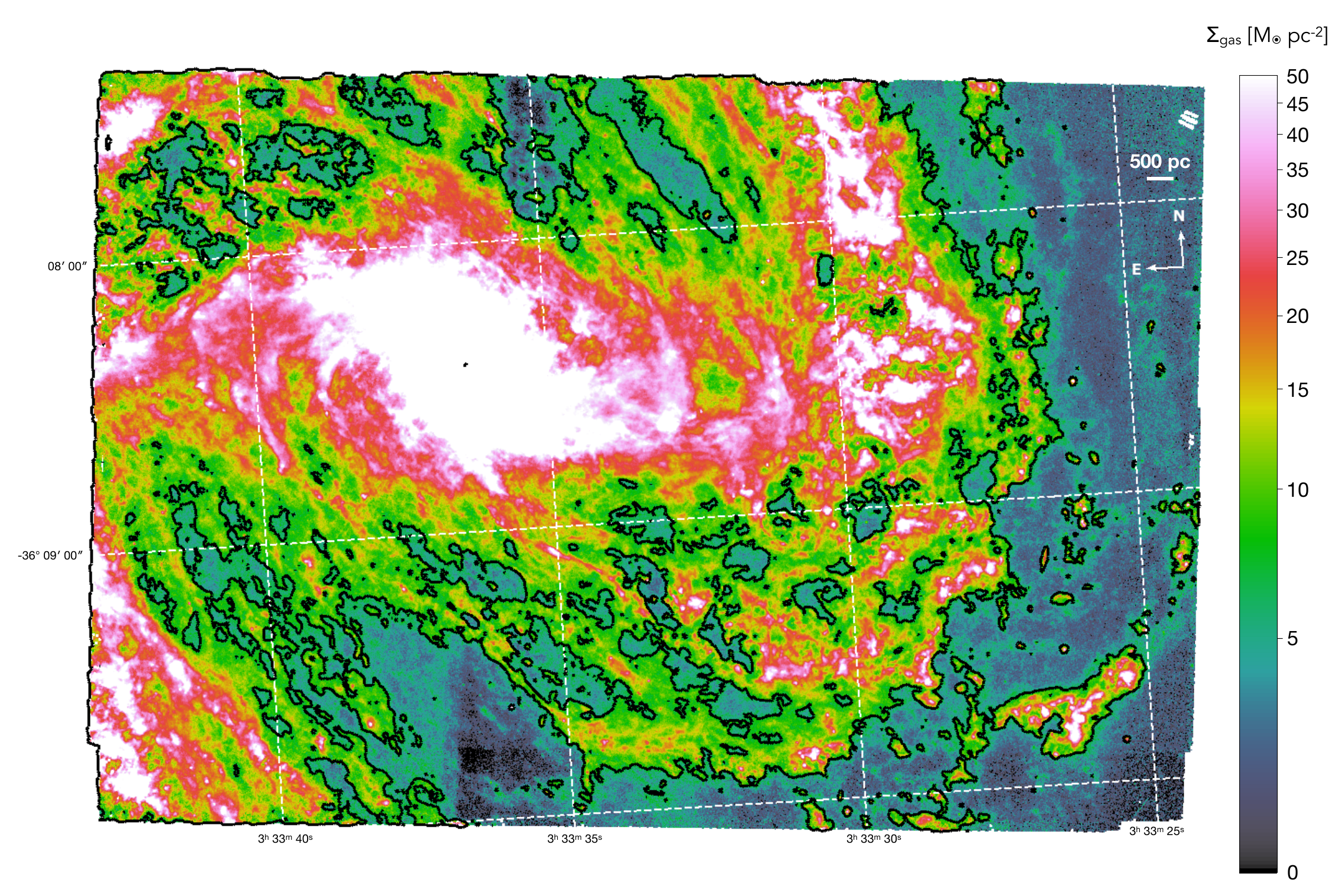}
    \caption{NGC~1365 gas surface density from scaled F1130W emission. A single black contour highlights the brightness level where the F1130W emission implies $\Sigma_{\rm gas} = 7$ \msun\ pc$^{-2}$. In NGC~1365, saturation in the center leads to an artifact where the columns of pixels with saturation show low values. This is evident in the faint regions of our map.}
    \label{fig:contour_1365}
\end{figure*}

\begin{figure*}
    \centering
    \includegraphics[height=12cm]{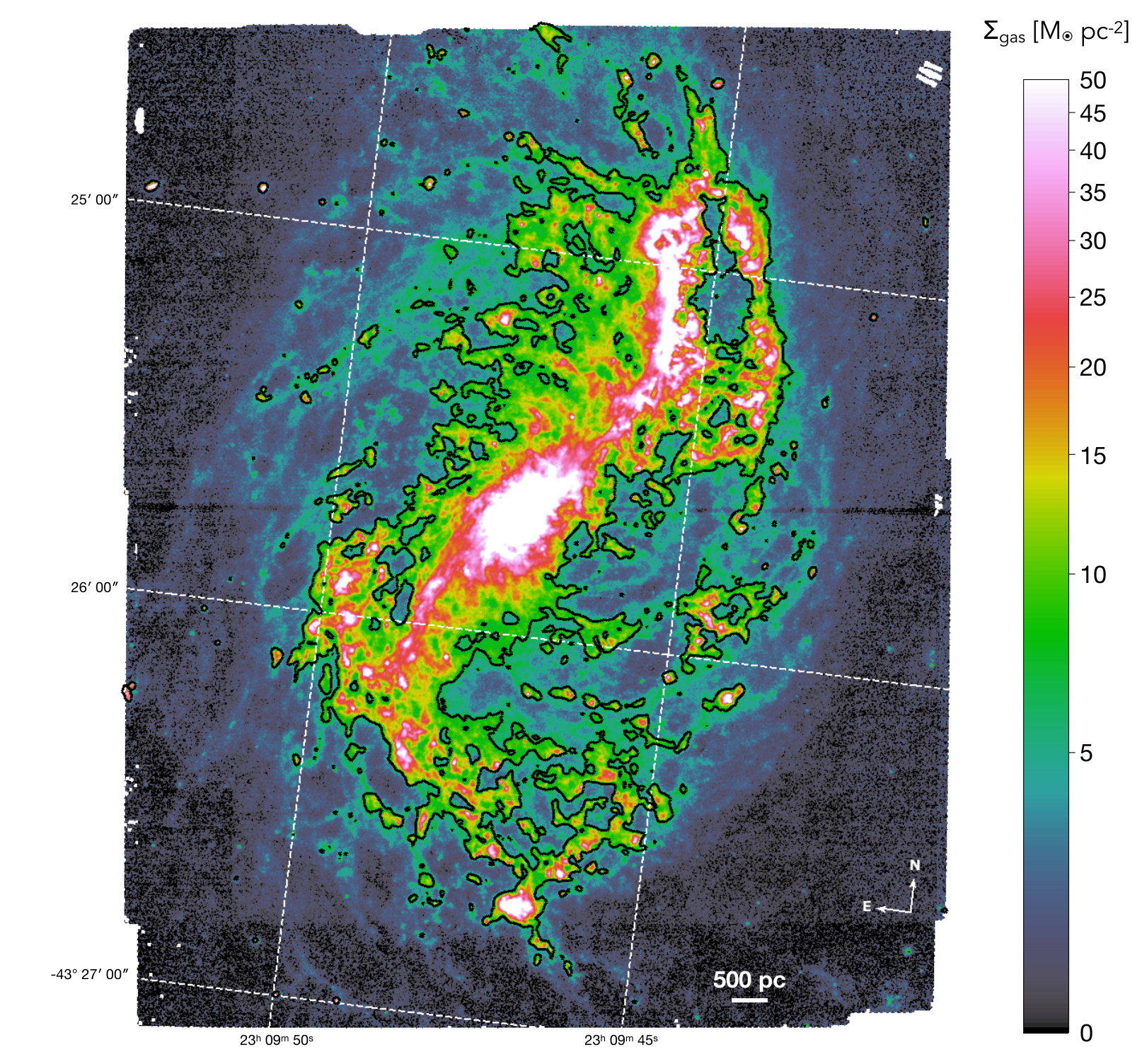}
    \caption{NGC~7496 gas surface density from scaled F1130W emission. A single black contour highlights the brightness level where the F1130W emission implies $\Sigma_{\rm gas} = 7$ \msun\ pc$^{-2}$. In NGC~7496, saturation in the center leads to an artifact where the columns of pixels with saturation show low values. This is evident in the faint regions of our map.}
    \label{fig:contour_7496}
\end{figure*}

In Figures~\ref{fig:contour_ic5332}  to \ref{fig:contour_7496} we show the F1130W emission re-scaled to gas surface density using the scalings quoted above for each galaxy.  In regions without bright CO emission, the diffuse ISM traced by our scaled PAH maps may include a combination of: (i) CO-bright~\htwo\ missed due to the current CO map sensitivity, (ii) a CO-dark \htwo\ component, or (iii) a mixture of atomic gas in the WNM and CNM phases. 

Shielding models for \htwo\ formation \citep{KrumholzMcKee2009ApJ...693..216K,SternbergLePetit2014ApJ...790...10S,BialySternberg2016ApJ...822...83B} predict the transition between \htwo\ or \hi\ dominated ISM at $\Sigma_{\rm HI} \approx 10\,(Z/Z_{\odot})^{-1}$~\msolpcsq. In the models, the transition point varies with the assumed geometry and role of self-shielding. In simulations which take into account the filamentary, turbulent nature of the gas, the column density of the \hi/\htwo\ transition can vary significantly \citep[e.g.][]{glover2007,glover2016}. Transitions at $\sim10$~\msolpcsq\ have been measured in the Milky Way \citep{StanimirovicMurray2014ApJ...793..132S,LeeStanimirovic2015ApJ...809...56L}. This transition is found at higher surface density in the Magellanic Clouds \citep{JamesonBolatto2016ApJ...825...12J}, following the expected trend with metallicity. Similarly, \citet{SchrubaBialy2018ApJ...862..110S} find a $\propto (Z/Z_{\odot})^{-1}$ dependence sampling a range of metallicities in 70 nearby galaxies, primarily on $>100$~pc scales.

The galaxies we target are all at approximately Solar metallicity, aside from IC~5332 which has a gradient extending down to $Z\sim 0.5 Z_\odot$. We expect therefore that \hi/H$_2$ transitions should occur above the noise floor of our maps ($3\sigma \sim 2$ \msolpcsq), suggesting that much of the faint emission we detect should be arising from atomic gas.  We select a threshold of $\Sigma_{\rm gas} = 7$ \msun\ pc$^{-2}$ as a conservative choice for where the gas should become atomic dominated given the predictions from shielding models at the relevant metallicities. The black contour on each image corresponds to this representative value of $\Sigma_{\rm gas} = 7$ \msun\ pc$^{-2}$. 

In the four galaxies we see a variety of behavior at low gas surface densities.  The maps of NGC~628 and NGC~1365 cover only parts of the galaxy disks, and the majority of pixels in these two maps have surface densities above 7 \msolpcsq.  In NGC~7496, the map extends into the outskirts of the disk, so we can observe the faint, low surface density structures in the outskirts.  IC~5332 has low surface density throughout its disk, and a large fraction of the emission in the map is below $7$ \msun\ pc$^{-2}$.  This is in good agreement with the faint CO emission observed in this galaxy \citep{leroy2021}. 

Many of the faint structures in all of the targets appear filamentary. Filaments in the PAH emission are explored in much greater detail in \citet{THILKER_PHANGSJWST} and \citet{MEIDT_PHANGSJWST} in this Issue.  In some cases, filamentary structures are observed to have widths comparable to a few resolution elements, and lengths that are substantially longer. For instance, in a few cases in NGC~628, we measure lengths of up to 400 pc for filamentary structures with $\Sigma_{\rm gas} < 7$ \msun\ pc$^{-2}$, yielding a 10:1 aspect ratio. Many of these filamentary structures appear as spurs or low surface density continuations of spiral arm structures. While such morphology may be expected as a consequence of the spiral structure of the galaxies, previous \hi\ observations have not had the resolution to characterize these outside the Local Group.  Throughout the diffuse ISM, structure is evident down to the noise level of the maps ($3\sigma \sim 2$ \msolpcsq).

At present, comparing our predicted surface densities from scaled F1130W to measured CO and \hi\ is limited by the available gas observations. The ALMA CO observations from the PHANGS survey have noise limits near this surface density \citep{leroy2021,LEROY1_PHANGSJWST}.  More limiting is the availability and resolution of \hi\ data. NGC~628 was observed as part of the THINGS survey \citep{walter2008} and the ``robust'' weighted map has $\sim6''$ resolution. NGC~7496 has recent MeerKAT observations at $~12''$ resolution. 
Future observations with higher resolution, higher S/N \hi\ and CO maps for the PHANGS-JWST targets will be necessary to measure gas surface densities for comparison to the F1130W emission at these faint levels.

\section{Measuring Diffuse ISM Topology with the Genus Statistic}\label{sec:genus}

One of the distinctive features of the faint regions of the F770W and F1130W maps is the clear presence of holes in the ISM distribution (see Figures~\ref{fig:contour_ic5332}  to \ref{fig:contour_7496}).  These holes may be due to dynamical effects or feedback, topics explored in more detail in \citet{WATKINS_PHANGSJWST}, \citet{BARNES_PHANGSJWST}, and \citet{MEIDT_PHANGSJWST}.  The pervasive emission in the PAH-tracing bands is one of the reasons these holes appear so clearly in the JWST maps. 
To quantify the topology of the faint emission, we use the {\em genus} statistic.
The genus statistic is the difference between the number of connected regions above and below a brightness threshold. A positive genus corresponds to a topology reflecting compact regions or sources that are bright compared to the threshold, surrounded by connected networks of lower brightness regions. Positive genus topology is referred to as source-dominated or a ``meatball''-like.  A negative genus is characterized by connected networks of emission above the threshold level and isolated holes below the threshold, which are referred to as a hole-dominated or a ``Swiss-cheese'' topology \citep[][]{Coles1988MNRAS.234..509C}.
By computing the genus through a range of intensity values, the resulting genus curve characterizes variations in the topology. When the genus curve crosses zero, the topology changes from ``Swiss-cheese'' to ``meatball''. 
The genus statistic is a well-known measure of spatial topology that was developed to study the topology of the cosmic microwave background \citep[][]{Coles1988MNRAS.234..509C} and has since been used to quantify the topology of the turbulent ISM, both in simulations and observations \citep[e.g., ][]{KowalLazarian2007ApJ...666L..69K,ChepurnovGordon2008ApJ...688.1021C,BurkhartStanimirovic2010ApJ...708.1204B,KochWard2017MNRAS.471.1506K,Burkhart2021PASP..133j2001B}.

Here we use the implementation of the Genus statistic in the \texttt{TurbuStat}\footnote{v1.3; \url{turbustat.readthedocs.io}} package \citep{Koch2019}.
We compute the genus curve for each scaled F1130W image at native resolution using a set of intensities chosen by percentiles in the images.
We discard all regions smaller than 20~pixels, $4\times$ the F1130W PSF area, in the images to minimize the effect of noise on the genus curves, where a connected region includes pixels adjacent along both the edges and corners.
We note that the zero-crossing point is not strongly affected by the choice of minimum area (for $10\mbox{--}40$~pixels).
There is no restriction on the morphology of a region (e.g., round vs. elongated).

We additionally impose three masking constraints on the data when computing the genus:
\begin{enumerate}
    \item We mask the PSF artifacts from the active galactic nuclei in NGC~1365 and NGC~7496. We use PSFs generated with WebbPSF and impose a masking cut-off at $1/1000^{\rm th}$ the maximum of the the normalized PSF.
    \item Two negative strips remain in the NGC~1365 (vertical through center, see Figure~\ref{fig:contour_1365}) and NGC~7496 (horizontal through center, see Figure~\ref{fig:contour_7496}) mosaics after our calibration. We impose by-hand masks to remove the visible negative regions.
    \item Finally, we apply a radial mask at $R_{\rm gal} > 0.4R_{25}$ to provide a consistent basis for comparison between the 4 targets. This corresponds to radial cut-offs of 3.2, 5.7, 13.7, 3.6 kpc in IC~5332, NGC~628, NGC~1365 and NGC~7496, respectively.
    This cut-off was chosen based on the range of $R_{\rm gal}$ that is well-sampled across all four galaxies, and where diffuse PAH emission is visually obvious in Figures \ref{fig:contour_ic5332} to \ref{fig:contour_7496}.
\end{enumerate}
Masks applied to the data will reduce the value of the genus statistic, roughly proportional to the total area, but will not change the relative shape within a single closed contour. 

Because of the wide dynamic range, we empirically choose 600~thresholds linearly sampled from the $0.01$ to $98^{\rm th}$~percentiles, and 50 logarithmically sampled thresholds from the $98^{\rm th}$ to $99.99^{\rm th}$~percentiles.
This ensures we sample the genus curve well near zero-crossings, tracing transitions from ``Swiss cheese'' to ``meatball'' topologies in the faint emission tail, while still characterizing the high-end intensity tail.

\begin{figure}[ht!]
\includegraphics[height=7cm]{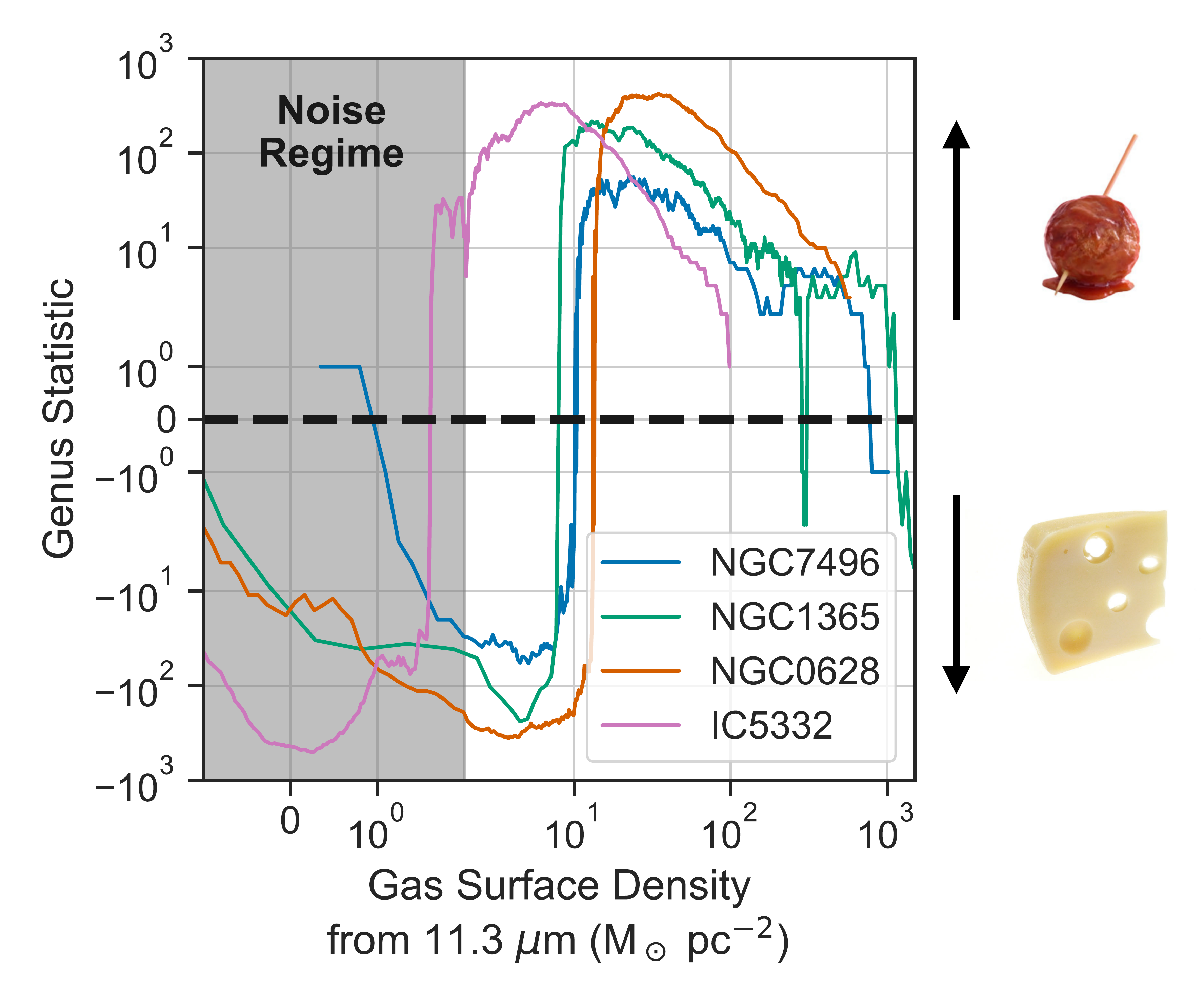}
\caption{Genus curves for the F1130W band, scaled to gas surface density (Equation~\ref{eq:ratio113}).
The gray region shows where the noise dominates the statistic up to $3\sigma$ significance above the background which corresponds to $\Sigma_{\rm gas} \approx 2$~\msolpcsq.
For a consistent comparison, we impose a radial cut at $0.4\,R_{25}$ in each galaxy, corresponding to 3.2, 5.7, 13.7, 3.6 kpc in IC 5332, NGC 628, NGC 1365 and NGC 7496, respectively.
The $11.3$~$\mu$m emission traces a clear conversion from a ``Swiss cheese'' (Genus$<0$) to ``meatball'' topology (Genus$>0$) when going from low to high gas surface density.
}
\label{fig:genus_1130}
\end{figure}

Figure~\ref{fig:genus_1130} shows the genus curves in the F1130W band\footnote{We note the F770W band has a similar topology and the genus analysis gives similar results, though with lower surface density sensitivity (see \S\ref{sec:obs}).} computed with a cut $R_{\rm gal} \, < \, 0.4R_{25}$.
There is a clear transition in the topology traced by the $11.3$~$\mu$m emission that varies by galaxy.
Table \ref{tab:genus_crossings} provides the zero crossing points from the genus curves.
The genus crossing corresponds to a gas surface density of $\sim8\mbox{--}10$~\msolpcsq \ (scaled from Eq. \ref{eq:ratio113}) for NGC~628, NGC~1365, and NGC~7496.
This closely matches the expected H$_2$-{\sc HI} transition point of $\sim10$~\msolpcsq\ based on shielding models at solar metallicity. 

The genus zero crossing point in IC~5332 is substantially lower, at $1.6$~\msolpcsq\ (0.3~MJy sr$^{-1}$), $\sim2\sigma$ significance above the background level of the map, and thus affected by noise fluctuations altering the genus curve.
A field of Gaussian noise will produce a similar transition from negative to positive genus, with a zero-crossing at the mean field value \citep{Coles1988MNRAS.234..509C}.
The significant detected structure alters the shape of the genus, as we see in Figure \ref{fig:genus_1130}, but for the sensitivity-limited IC~5332, the genus zero-crossing is qualitatively consistent with a noise-like field.
The diffuse PAH emission in IC~5332 being sensitivity-limited is a substantially lower filling fraction in the PAH emission compared to the other three galaxies (Figure \ref{fig:contour_ic5332}).
Given that IC~5332 is the only dwarf spiral in this sample, with a lower stellar mass, star formation rate, and metallicity relative to the other galaxies, this lower filling fraction at a common sensitivity limit is expected \citep[for example,][in the Magellenic Clouds]{Jameson2018,Chevance2020b}.
The genus zero crossing point for IC~5332 then indicates that the genus is limited to measuring topology changes for tracers with a high filling fraction (see \S\ref{sec:discussion_topology}).

\begin{deluxetable}{l|c|c}[t!]
\tabletypesize{\small}
\tablecaption{Genus zero crossings \label{tab:genus_crossings}}
\tablewidth{0pt}
\tablehead{
\colhead{Target} & 
\colhead{Intensity} &
\colhead{$\Sigma_{\rm gas}$} \\
\colhead{} & 
\colhead{(MJy sr$^{-1}$)} &
\colhead{(M$_\odot$ pc$^{-2}$)}
}
\startdata
IC 5332  &  0.3  & 1.6 \\
NGC 628 &  2.5  & 13.0 \\  
NGC 1365 &  1.5 & 8.0 \\
NGC 7496 &  1.9  & 10.1 
\enddata
\tablecomments{Zero crossings computed from the genus curves in Figure \ref{fig:genus_1130}. We scale to a gas surface density using Eq. \ref{eq:sdscalef1130w}.}
\end{deluxetable}

\section{Discussion} \label{sec:disc}

\subsection{Morphology of the Faint PAH Emission}

In the regions of our F1130W maps corresponding to gas surface densities $<7$ \msolpcsq, we see a variety of structures down to our detection limit of  $\sim2$ \msolpcsq, including filaments, clumps, bubbles, inter-arm spurs, and faint structures in the outskirts of galaxies.  The 10--40 pc scales of PAH emission and gas column we explore here are comparable only to previous observations of the Milky Way and Local Group. In the Milky Way, the 8~$\mu$m Spitzer/IRAC map from GLIMPSE similarly showed filamentary structures, pervasive PAH emission from diffuse gas, and a porous, bubble-filled structure for the ISM.  PAH emission in the Milky Way is observed to be associated with both molecular and atomic gas phases \citep{ChurchwellBabler2009PASP..121..213C}. 

The external perspective provided by nearby galaxies at JWST's resolution may also enable identification of filamentary structures in both molecular and atomic-dominated gas. For example, at 10${-}$40 pc resolution, our maps reveal filamentary structures with aspect ratios of 10:1 or more and lengths of several 100 pc in both interarm regions (e.g.\ NGC~628) and in the outskirts of galaxies \citep[NGC~7496, IC~5332; see also][]{THILKER_PHANGSJWST, MEIDT_PHANGSJWST}. While these filaments are wider ($>10-40$ pc) and likely associated with diffuse atomic gas, the filament aspect ratios are similar to giant molecular filaments identified in the Milky Way \citep{ragan2014,zucker2018,hacar2022}. These may be similar to giant \hi\ filaments \citep{SyedSoler2022A&A...657A...1S}, which have proven difficult to identify even in the Milky Way due to line-of-sight confusion \citep{SolerBeuther2020A&A...642A.163S}.

One exciting possibility with gas tracers at 10-40 pc scales extending to \hi\ dominated gas is structurally separating the CNM and WNM. The CNM is dense atomic gas where H$_2$ formation proceeds, and therefore the balance between the WNM and CNM is central to the regulation of star formation.  With existing \hi\ observations, separation of the WNM and CNM outside the Local Group relies on Gaussian decomposition of \hi\ emission spectra, which is uncertain at low spatial resolution \citep{koch2021}.
On $\sim10$ pc scales, the WNM and CNM are expected to have distinct spatial structures---Galactic studies separating the WNM and CNM phases in the Milky Way primarily find a diffuse widespread WNM component and small-scale CNM filamentary networks \citep[e.g., ][]{ClarkPeek2014ApJ...789...82C,smith2014,KalberlaHaud2018A&A...619A..58K,MarchalMiville-Deschenes2019A&A...626A.101M,kalberla2021}.
This morphological separation is limited by the \hi\ resolution and sensitivity restricting these studies to the Milky Way, but early GASKAP work points to recovering a pervasive filamentary structure on $10$~pc scales in the Magellanic Clouds \citep{PingelDempsey2022PASA...39....5P}. 
Given that different phases of the neutral medium have different morphologies, the PAH emission map could allow morphological separation of CNM/WNM components. The emission morphology sets its topology, so the genus statistic and related analyses may offer a new way to characterize ISM phase structure.

\subsection{The Transition in ISM Topology at $\sim10$ \msolpcsq}
\label{sec:discussion_topology}

A striking visual impression of the JWST F1130W (and other PAH tracing maps) is the clear presence of bubbles throughout the ISM \citep[see][this issue]{WATKINS_PHANGSJWST,BARNES_PHANGSJWST}. These bubbles are clearly visible because they occur in a pervasive background of PAH emission throughout the galaxies.  The genus statistic allows a quantification of this visually evident topology.  At faint levels in the maps, the genus statistic is negative, indicating more closed structures below the threshold than above, or a ``Swiss cheese''-like topology. From our analysis, we found that the PAH emission transitions from ``Swiss cheese'' to ``meatball''-like topology (i.e.\ sources rather than holes) at  $\sim10$ \msolpcsq\ in all galaxies where the topology transition is well measured (NGC~628, 1365, and 7496). Interestingly, this threshold is close to what is expected for the \hi/H$_2$ transition.

Our observation of the topology transition point matches well with the expectation that the pervasive, relatively smooth gas distribution in which holes and bubbles are evident is atomic-dominated. Previous observations of nearby galaxies have shown that atomic and molecular gas show distinct spatial distributions.  Molecular gas observed in CO emission is found to be ``clumpy,'' whereas atomic gas shows smoother distributions as traced by 21-cm \hi\ emission on the $>100$~pc scales that are representative of the WNM \citep{leroy2013}. However, part of this contrast in structure stems from observational limitations of the 21-cm \hi\ resolution and the CO sensitivity, and so represents the largest possible difference in topologies between the phases.
Physically one might also expect this topology transition at $\sim10$ \msolpcsq\ if it is also related to gravitational or thermal instability above this column causing gas to collapse into clumps \citep{dobbs2014}, which the genus statistic should measure through a transition in topology.

\begin{figure*}
    \centering
    \plotone{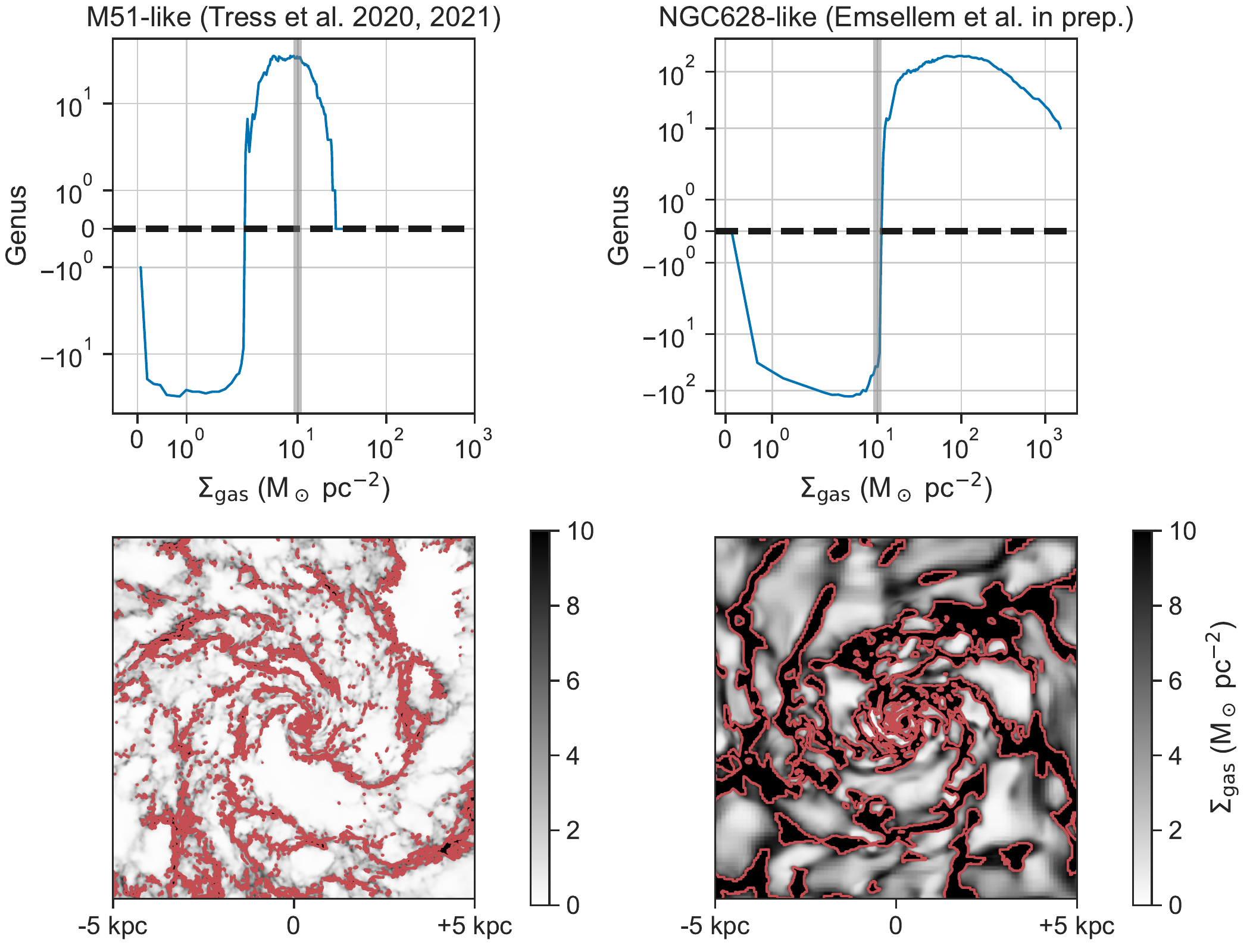}
    \caption{
    Genus curves (top panels) from the gas surface density maps for two simulation snapshots (bottom panels).
    We show a gray vertical line at $\Sigma_{\rm gas}=10$~\msolpcsq as visual reference point to compare the two genus curves. 
    The red contour indicates the surface density at the topology change measured by the genus zero crossing point. \textbf{Left:} M51-like simulation \citep{Tress2020,Tress2021}. \textbf{Right:} NGC628-like simulation (Emsellem et al. in prep.).}
    \label{fig:sims_genus}
\end{figure*}

\subsection{Comparison to Simulations}
High-resolution ISM column density maps from PAHs open new opportunities for comparing ISM topology between observations and simulations \citep{ChepurnovGordon2008ApJ...688.1021C,BurkhartStanimirovic2010ApJ...708.1204B,KochWard2017MNRAS.471.1506K}.
We briefly explore this application with the genus statistic by comparing between two total neutral gas surface density maps of simulated galaxies, one from an M\,51-like galaxy \citep{Tress2020,Tress2021} and an NGC\,628-like galaxy (Emsellem et al. in prep.).
Briefly, the NGC\,628-like simulation was initialised using known properties of NGC\,628 in terms of stellar mass, gas content and distribution and viewing angles, and run with the adaptive-mesh refinement {\sc Ramses} code \citep{teyssier2002} at a maximum sampling of about 3.7~parsec: it includes atomic and molecular cooling, a UV radiation background, star formation with a fixed efficiency per free-fall time above a given density threshold, and feedback from type Ia and type II SNe, stellar winds and radiation pressure (more details in Emsellem et al. in prep).
The M51-like simulation was initialized to have properties similar to M\,51. \citet{Tress2020,Tress2021} ran simulations both with and without an interacting companion, but here we just consider the simulation without a companion, in which the galaxy develops into a flocculent spiral. The simulation was run with the {\sc arepo} moving mesh code \citep{springel2010} and achieves a maximum spatial resolution of around 0.5~parsec. It includes a similar range of physical processes to the {\sc Ramses} calculation, but does not include stellar winds or radiation pressure. Further details can be found in \citet{Tress2020}. Figure \ref{fig:sims_genus} shows the genus curves and gas surface density maps of the two snapshots, with the contour at the genus zero-crossing point.

We recover the two behaviors we observed using the scaled PAH emission in Figure \ref{fig:genus_1130}: (i) the M51-like simulation has a lower filling fraction of diffuse gas that leads to a lower genus crossing point around $\Sigma_{\rm gas} \sim 2$~\msolpcsq, 
(ii) the NGC628-like simulation has a broad diffuse component and a genus crossing point at $\Sigma_{\rm gas} \sim 10$~\msolpcsq, remarkably similar to the transition we find for NGC628, 1365, and 7496.
These crossing points remain at similar surface densities after varying the resolution of the two simulated maps to match the resolution of the JWST-MIRI data.
However, more work remains to map from variations in the ISM morphology to the underlying physics contained in the simulations.  The small transition value for the M51-like simulation differs clearly from our observations and the other simulations. However, this simulation reaches higher resolution than the observations (sub-pc in dense regions) and the contour structures appear notably different in Figure \ref{fig:sims_genus}.  This difference suggests that this approach would require testing the effects of changing resolution. We will be able to test this effect and others further by making similar comparisons using a suite of simulations and the complete PHANGS-JWST sample.

Methods like the genus provide a promising route to improve our interpretation of the neutral ISM morphology at high resolution in both observations and simulations: the high-resolution JWST PAH maps provide new constraints for simulations to reproduce the low surface density morphology, while close correspondence with topology measures in simulations can help us better interpret how the topology varies across the molecular to atomic ISM phases in observations.

\section{Conclusions} \label{sec:conc}

In this Letter, we present a test case for using PAH emission observed with JWST-MIRI as a high resolution, high sensitivity, cold ISM phase-agnostic, tracer of gas. We use scalings determined in H$_2$-dominated (as traced by CO), but diffuse ISRF heated gas, and apply those to the faint PAH emission in the F1130W map. With these simplistic assumptions, we trace gas surface densities down to $\Sigma_{\rm gas} \sim 2$ \msolpcsq\ at $3\sigma$ at $10-40$\,pc resolution. This approach is enabled by the high angular resolution of JWST, which allows ISM gas outside the sphere of influence of star forming regions, heated by the diffuse ISRF, to be isolated to calibrate our gas tracer. 

Using these gas maps, we isolate regions at $<7$ \msolpcsq\ which we expect to be atomic gas dominated. The existing \hi\ and CO maps are not sensitive or high enough resolution to determine definitively if the faint PAH emission arises from atomic gas, CO-bright H$_2$ below the detection threshold, or CO-dark H$_2$. Given the predictions from H$_2$ formation models and the $\sim$Solar metallicity of our targets, we expect that gas at $<7$ \msolpcsq\ is primarily atomic.

At these low surface densities, we see a wealth of structures, from filaments, clumps, inter-arm emission, bubbles, and faint galaxy outskirts.  These structures are inaccessible at this resolution in our targets with current \hi\ and CO observations.  We highlight the possibility of using PAH emission to separate the CNM and WNM gas by its spatial structure, which would make the WNM/CNM balance an accessible observable in galaxies outside the Local Group. We also note that the aspect ratios and lengths of filamentary structures evident at low gas surface densities in our maps suggest we may be seeing the equivalent of giant \hi\ filaments in other galaxies.  

We also measure the topology of the PAH emission using the genus statistic.  We find that the PAH emission at faint levels shows a ``Swiss cheese'' like morphology, suggesting relatively smooth, diffuse emission with holes and bubbles.  At around 10 \msolpcsq\ we find a transition from ``Swiss cheese''-like to ``meatball''-like or source dominated topology.  The fact that this transition occurs at a similar surface density to the \hi/H$_2$ transition agrees with previous observations that H$_2$ tends to have a clumpy distribution while \hi\ is smoother.

This initial investigation shows the potential for using JWST observations of PAH emission to reveal key aspects of the structure and phase of ISM gas in nearby galaxies. Future work that calibrates these relationships with higher resolution and sensitivity \hi\ and CO will be necessary. In addition, deeper understanding of the variation of dust-to-gas ratios, PAH fractions, and ISRF intensity and spectra will be crucial to fully making use of this new tool.  

\section*{Acknowledgements}
We thank the anonymous referee for helpful feedback that improved the paper.
This work is based on observations made with the NASA/ESA/CSA James Webb Space Telescope. The data were obtained from the Mikulski Archive for Space Telescopes at the Space Telescope Science Institute, which is operated by the Association of Universities for Research in Astronomy, Inc., under NASA contract NAS 5-03127 for JWST. These observations are associated with program 2107. 
The specific observations analyzed can be accessed via \dataset[10.17909/9bdf-jn24]{http://dx.doi.org/10.17909/9bdf-jn24}

KS acknowledges funding support from grant support by JWST-GO-02107.006-A.
EWK acknowledges support from the Smithsonian Institution as a Submillimeter Array (SMA) Fellow and the Natural Sciences and Engineering Research Council of Canada.
HAP acknowledges support by the National Science and Technology Council of Taiwan under grant 110-2112-M-032-020-MY3
JMDK gratefully acknowledges funding from the European Research Council (ERC) under the European Union's Horizon 2020 research and innovation programme via the ERC Starting Grant MUSTANG (grant agreement number 714907). COOL Research DAO is a Decentralized Autonomous Organization supporting research in astrophysics aimed at uncovering our cosmic origins.
MB acknowledges support from FONDECYT regular grant 1211000 and by the ANID BASAL project FB210003.
IC thanks the National Science and Technology Council for support through grants 108-2112-M-001-007-MY3 and 111-2112-M-001-038-MY3, and the Academia Sinica for Investigator Award AS-IA-109-M02.
EJW acknowledges the funding provided by the Deutsche Forschungsgemeinschaft (DFG, German Research Foundation) -- Project-ID 138713538 -- SFB 881 (``The Milky Way System'', subproject P1). 
KK gratefully acknowledges funding from the Deutsche Forschungsgemeinschaft (DFG, German Research Foundation) in the form of an Emmy Noether Research Group (grant number KR4598/2-1, PI Kreckel).
MC gratefully acknowledges funding from the DFG through an Emmy Noether Research Group (grant number CH2137/1-1).
TGW and ES acknowledge funding from the European Research Council (ERC) under the European Union’s Horizon 2020 research and innovation programme (grant agreement No. 694343).
FB would like to acknowledge funding from the European Research Council (ERC) under the European Union’s Horizon 2020 research and innovation programme (grant agreement No.726384/Empire).
RSK acknowledges financial support from the European Research Council via the ERC Synergy Grant ``ECOGAL'' (project ID 855130), from the Deutsche Forschungsgemeinschaft (DFG) via the Collaborative Research Center ``The Milky Way System''  (SFB 881 -- funding ID 138713538 -- subprojects A1, B1, B2 and B8) and from the Heidelberg Cluster of Excellence (EXC 2181 - 390900948) ``STRUCTURES'', funded by the German Excellence Strategy. RSK also thanks the German Ministry for Economic Affairs and Climate Action for funding in  project ``MAINN'' (funding ID 50OO2206). 
MQ acknowledges support from the Spanish grant PID2019-106027GA-C44, funded by MCIN/AEI/10.13039/501100011033.
ER and HH acknowledge the support of the Natural Sciences and Engineering Research Council of Canada (NSERC), funding reference number RGPIN-2022-03499.
This research was supported by the Excellence Cluster ORIGINS which is funded by the Deutsche Forschungsgemeinschaft (DFG, German Research Foundation) under Germany's Excellence Strategy - EXC-2094-390783311. Some of the simulations in this paper have been carried out on the computing facilities of the Computational Center for Particle and Astrophysics (C2PAP). EE would like to thank Alexey Krukau and Margarita Petkova for their support through C2PAP.
KG is supported by the Australian Research Council through the Discovery Early Career Researcher Award (DECRA) Fellowship DE220100766 funded by the Australian Government. 
KG is supported by the Australian Research Council Centre of Excellence for All Sky Astrophysics in 3 Dimensions (ASTRO~3D), through project number CE170100013. 
JK gratefully acknowledges funding from the Deutsche Forschungsgemeinschaft (DFG, German Research Foundation) through the DFG Sachbeihilfe (grant number KR4801/2-1).
AS is supported by an NSF Astronomy and Astrophysics Postdoctoral Fellowship under award AST-1903834.
AKL gratefully acknowledges support by grants 1653300 and 2205628 from the National Science Foundation, by award JWST-GO-02107.009-A, and by a Humboldt Research Award from the Alexander von Humboldt Foundation.
G.A.B. acknowledges the support from ANID Basal project FB210003.

\vspace{5mm}
\facilities{JWST, ALMA}

\software{astropy \citep{2013A&A...558A..33A,2018AJ....156..123A,AstropyCollaboration2022ApJ...935..167A}, TurbuStat \citep{Koch2019}, WebbPSF \citep{Perrin2014SPIE.9143E..3XP}}

\bibliography{pahmaps.bbl}
\bibliographystyle{aasjournal}

\suppressAffiliationsfalse
\allauthors

\end{document}